\documentclass[draft]{eptcs}

\usepackage{etoolbox}
\newtoggle{proofs}
\togglefalse{proofs}    % only in eptcs style

\usepackage{amsthm}

\usepackage{enumerate}
\usepackage{mathtools} 
\usepackage{amsmath}
\usepackage{amssymb}
\usepackage{eucal}
\usepackage{tikz}
\usepackage{cases} % must come after amsmath
\usepackage{environ, thmtools, thm-restate}  % Advanced theorem handling

\usepackage{stmaryrd}
\usepackage{graphicx}
\usepackage{nicefrac}
\usepackage{DotArrow}

\usepackage{paralist} % for inlined itemize

\usepackage[nomargin,inline,index]{fixme} % Simplified management of FIXME's
\fxusetheme{color}
\usepackage{ifdraft}                % Introduces \ifdraft

\usepackage{hyperref} % Load cleveref last!
\hypersetup{final} % Needed to generate hyperlinks even in draft mode
\usepackage{cleveref}

\usepackage{appendix}

\FXRegisterAuthor{tizi}{antizi}{\color{blue} {\underline{tizi}}}
\FXRegisterAuthor{bart}{anbart}{\color{magenta} {\underline{bart}}}
\FXRegisterAuthor{michele}{anmichele}{\color{red} {\underline{michele}}}
\FXRegisterAuthor{paolo}{anpaolo}{\color{cyan} {\underline{paolo}}}
\FXRegisterAuthor{zun}{anzun}{\color{green} {\underline{Zun}}}

\newcommand{\hidden}[1]{}

\newcommand{\irule}[2]{\dfrac{#1}{#2}}
\declaretheorem[name=Theorem]{theorem}

\declaretheorem[name=Lemma,numberlike=theorem]{lemma}

\declaretheorem[name=Definition,numberlike=theorem]{definition}

\declaretheorem[name=Corollary,numberlike=theorem]{corollary}

\newcommand{\powset}[1]{\wp(#1)}

\newcommand{\subseteqfin}{\subseteq_{\it fin}}

\newcommand{\seq}[1]{\langle {#1} \rangle}
\renewcommand{\epsilon}{\varepsilon}
\newcommand{\mkset}[1]{\overline{#1}}

\newcommand{\bind}[2]{\nicefrac{#2}{#1}}
\newcommand{\setenum}[1]{\{#1\}}
\newcommand{\setcomp}[2]{\{{#1} \mid {#2}\}}

\newcommand{\ES}{\textrm{\textup{ES}}}

\newcommand{\esname}{\mathcal{E}}

\newcommand{\Ev}{\mathbf{E}}
\newcommand{\Act}{\mathbf{A}}
\newcommand{\aname}{\mathbf{P}}
\newcommand{\coAct}{\overline{\Act\!}\,}
\newcommand{\act}[1]{\mathit{{#1}}}
\newcommand{\coact}[1]{\overline{\act{#1}}}
\newcommand{\success}[1][]{\mathbf{1}_{{#1}}}
\newcommand{\mmid}{\,\|\,}

\newcommand{\CF}[1]{{\it CF}{({#1})}}

\newcommand{\event}[2]{{\pmv #1}:{\atom #2}}

\newcommand{\cname}{\mathcal{C}}

\newcommand{\princsym}{\pi}
\newcommand{\princ}[2][]{\princsym_{#1}({#2})}
\newcommand{\invprinc}[2][]{\Ev_{#2}}

\newcommand{\winsym}{\mathcal{W}}
\newcommand{\win}[1][]{\winsym_{#1}}

\newcommand{\payoffsym}[1][]{\Phi_{#1}}
\newcommand{\payoff}[1][]{\payoffsym[#1]}

\newcommand{\ccomp}[0]{\mid}

\newcommand{\pmv}[1]{\ensuremath{\mathit{#1}}}

\newcommand{\atom}[1]{\textit{#1}}

\newcommand{\sep}{\ \bnfmid\ }

\newcommand{\nil}{\mathbf{0}}

\newcommand{\bnfmid}{\;\big|\;}

\newcommand{\sem}[3][]{\mbox{\ensuremath{\llbracket#2\rrbracket_{#3}^{#1}}}}
\newcommand{\SumIntRaw}{\bigoplus}
\newcommand{\SumExtRaw}{\sum}

\newcommand{\sumInt}{\oplus}
\newcommand{\sumExt}{+}

\newcommand{\SumInt}[3][]{\SumIntRaw_{#1} {#2} \, . \, {#3}}
\newcommand{\SumExt}[3][]{\SumExtRaw_{#1} {#2} \, . \, {#3}}

\newcommand{\sumI}[2]{{#1} \, . \, {#2}}
\newcommand{\sumE}[2]{{#1} \, . \, {#2}}

\newcommand{\rec}[2]{\mathit{rec}\;{#1}.\;{#2}}

\newcommand{\cmove}[1]{\xrightarrow{#1}\hspace{-1.8ex}\rightarrow}

\makeatletter
\newcommand*{\da@rightarrow}{\mathchar"0\hexnumber@\symAMSa 4B }
\newcommand*{\da@leftarrow}{\mathchar"0\hexnumber@\symAMSa 4C }
\newcommand*{\xdashrightarrow}[2][]{%
  \mathrel{%
    \mathpalette{\da@xarrow{#1}{#2}{}\da@rightarrow{\,}{}}{}%
  }%
}
\newcommand{\xdashleftarrow}[2][]{%
  \mathrel{%
    \mathpalette{\da@xarrow{#1}{#2}\da@leftarrow{}{}{\,}}{}%
  }%
}
\newcommand*{\da@xarrow}[7]{%
  \sbox0{$\ifx#7\scriptstyle\scriptscriptstyle\else\scriptstyle\fi#5#1#6\m@th$}%
  \sbox2{$\ifx#7\scriptstyle\scriptscriptstyle\else\scriptstyle\fi#5#2#6\m@th$}%
  \sbox4{$#7\dabar@\m@th$}%
  \dimen@=\wd0 %
  \ifdim\wd2 >\dimen@
    \dimen@=\wd2 %   
  \fi
  \count@=2 %
  \def\da@bars{\dabar@\dabar@}%
  \@whiledim\count@\wd4<\dimen@\do{%
    \advance\count@\@ne
    \expandafter\def\expandafter\da@bars\expandafter{%
      \da@bars
      \dabar@ 
    }%
  }%  
  \mathrel{#3}%
  \mathrel{%   
    \mathop{\da@bars}\limits
    \ifx\\#1\\%
    \else
      _{\copy0}%
    \fi
    \ifx\\#2\\%
    \else
      ^{\copy2}%
    \fi
  }%   
  \mathrel{#4}%
}
\makeatother

\newcommand{\TS}[1]{\mathsf{TS}(#1)}
\newcommand{\ETS}[1]{\mathsf{ES}(#1)}
\newcommand{\EvTS}[1]{\mathsf{EvS}(#1)}

\newcommand{\mytitle}{A note on two notions of compliance}
\title{\mytitle}

\author{Massimo Bartoletti
\institute{Dipartimento di Matematica e Informatica\\
University of Cagliari, Italy}
\email{bart@unica.it}
\and
Tiziana Cimoli
\institute{Dipartimento di Matematica e Informatica\\
University of Cagliari, Italy}
\email{t.cimoli@unica.it}
\and
G.~Michele Pinna
\institute{Dipartimento di Matematica e Informatica\\
University of Cagliari, Italy}
\email{gmpinna@unica.it}
}

\def\titlerunning{\mytitle}
\def\authorrunning{M. Bartoletti, T. Cimoli, G.M. Pinna}

\begin{document}

\maketitle

\begin{abstract}
We establish a relation between two models of contracts:
binary session types,
and a model based on event structures and game-theoretic notions.
In particular, we show that 
compliance in session types
corresponds to the existence of certain winning strategies
in game-based contracts.
\end{abstract}

\section{Introduction}

Several recent papers have been devoted to the study of \emph{contracts} as a
way to formally specify abstractions of the behaviour of software systems. 
A common aspect that gathers together some of these studies is a notion of \emph{compliance}. 
This is a relation between systems which want to interact. %with each other.
%Before running the actual systems, 
Before starting the interaction, %their 
contracts are statically checked for compliance:
when enjoyed, it guarantees that systems respecting their contracts will interact correctly. 
Since distributed applications are often constructed by dynamically discovering and
composing services published by different (possibly distrusting) organizations, compliance
becomes relevant to protect those services from each other's misbehaviour. 
Indeed, the larger an application is, the greater is the
probability that some of its components deviates from the expected
behaviour (either because of unintentional bugs, or maliciousness).

To obtain protection, compliance can be modelled in many different ways. 
Typically, it is formalised as a fairness property, which
ensures progress (possibly, until reaching a success state~\cite{Castagna09toplas,Barbanera10ppdp}), 
or which ensures the possibility of always reaching success from any state~\cite{Bravetti07fsen,Aalst10cj}. 
Weaker variants of compliance allow services to discard some messages~\cite{Barbanera14ice},
or involve orchestrators which can suitably rearrange them~\cite{Padovani10tcs}.

While all the above approaches express contracts as processes of some
process algebra, in~\cite{BCZ13post} contracts are modelled as multi-player
concurrent games, whose moves are transitions in an event structure~\cite{Winskel86},
and where compliance is defined as the existence of winning
strategies in these games. 
By abstracting away from the concrete details of process calculi, 
this model may be used as a unifying framework for reasoning about contracts, 
in the same spirit that event structures are used as 
an underlying semantics for a variety of concrete models of concurrency.

As a first step towards unifying different views of contracts, 
in this paper we interpret binary session types~\cite{Honda98esop}
as game-based contracts, 
by providing them with an event structure semantics 
(\Cref{def:es-denotational}).
Our main technical contribution is that compliance in the former model 
corresponds to the existence of a certain kind of winning strategies in the latter 
(\Cref{th:compliance-iff-eager-winning}).
The constructions used to obtain this result suggest that also other notions of compliance 
(\emph{e.g.}, \emph{I/O compliance}~\cite{BSZ14concur},
Padovani's \emph{weak compliance}~\cite{Padovani10tcs} and 
Barbanera \& de' Liguoro's \emph{\texttt{\textup{skp}}-compliance}~\cite{Barbanera14ice}) 
might be expressed game-theoretically by suitably adjusting the event structure semantics, 
and by restricting the class of admissible winning strategies.

\section{Session types}

Let $\Act$ be a set of \emph{actions}, 
ranged over by $\act{a}, \act{b}, \ldots$,
and let $\coAct = \setcomp{\coact{a}}{\act{a} \in \Act}$
be such that $\Act \cap \coAct = \emptyset$.
We let $\alpha,\beta,\ldots$ range over $\Act \cup \coAct$.
In~\Cref{def:sb-contracts:syntax} we introduce the syntax of 
\emph{binary} session types,
following the notation used in~\cite{Barbanera10ppdp}.

\begin{definition}[\bf Session type] \label{def:sb-contracts:syntax}
\emph{Session types} are defined as follows:
\begin{align*}
    P,Q \;\; & ::= \;\;
\textstyle
    \success
    \ \sep \ 
    \SumInt[i \in I]{\coact{a}_i}{P_i} 
    \ \sep \ 
    \SumExt[i \in I]{\act{a}_i}{P_i} 
    \ \sep \
    \rec{x}{P}
    \sep \ 
    x
\end{align*}
where % we assume that
$(i)$ the index set $I$ is finite and non-empty,
$(ii)$ the actions in internal/external choices are pairwise distinct, and
$(iii)$ recursion is guarded.
% A \emph{bilateral session type} is a pair $P \mmid Q$.
\end{definition}

Session types are processes of a process algebra
featuring $\success$ (success), 
internal choice $\SumInt[i \in I]{\coact{a}_i}{P_i}$,
external choice $\SumExt[i \in I]{\act{a}_i}{P_i}$,
and guarded recursion.
If $Q = \SumInt[i \in I]{\coact{a_i}}{P_i}$ and $0 \not\in I$,
we write $\coact{a_0}.P_0 \sumInt Q$ for $\SumInt[i \in I \cup \setenum{0}]{\coact{a_i}}{P_i}$
(same for external choice).

\begin{figure}[t]
\small
\hrulefill
\vspace{-5pt}
\[
\begin{array}{ll}
  \sumI{\coact{a}}{P} \,\sumInt\, Q
  \;\xrightarrow{}\;
  \sumI{\coact{a}}{P}
  \hspace{20pt}
  &
  \sumI{\coact{a}}{P}
  \;\xrightarrow{\coact{a}}\;
  P
  \\[4pt]
  \sumE{\act{a}}{P} \,\sumExt\, Q
  \;\xrightarrow{\act{a}}\;
  P
  &
  \rec x P \xrightarrow{} P\setenum{\bind{x}{\rec x P}}
\end{array}
\hspace{20pt}
\begin{array}{c}
  \irule
  {P \xrightarrow{} P'}
  {P \mmid Q \xrightarrow{} P' \mmid Q}
  % \irule
  % {Q \xrightarrow{} Q'}
  % {P \mmid Q \xrightarrow{} P \mmid Q'}
  \hspace{25pt}
  \irule
  {P \xrightarrow{\coact{a}} P' \quad Q \xrightarrow{\act{a}} Q'}
  {P \mmid Q \xrightarrow{} P' \mmid Q'}
\end{array} 
\vspace{-5pt}
\]
\hrulefill
\vspace{-5pt}
\caption{Operational semantics of session types (symmetric rules omitted).}
\label{fig:sb-contracts:semantics}
\end{figure}

\medskip
The semantics of session types is defined
in~\Cref{fig:sb-contracts:semantics}.
The intuition is that 
a session type models the intended behaviour of \emph{one} 
of the two participants involved in a session,
while the behaviour of two interacting participants is modelled 
by the composition of two session types, denoted $P \mmid Q$.
An internal choice must first commit to one of the branches $\coact{a}.P$,
before advertising $\coact{a}$.
An external choice can always advertise each of its actions.
There, participants can run asynchronously
only when committing to a branch or unfolding recursion.
Synchronisation requires that a participant has committed to
a branch $\coact{a}$ in an internal choice, and the other
offers $\act{a}$ in an external choice.

Following~\cite{Laneve07concur,Castagna09toplas,Barbanera10ppdp}
we define a notion of compliance between session types.
The intuition is that if a client contract $P$ is compliant with 
a server contract $Q$ then,
whenever a computation of $P \mmid Q$ becomes stuck,
the client has reached the success state.

\begin{definition}[Compliance]
\label{def:compliance}
$P$ is \emph{compliant} with $Q$ (written $P \dashv Q$) iff
$P \mmid Q \rightarrow^* P' \mmid Q' \not\rightarrow\;$ implies $P' = \success$.
\end{definition}

\section{Contracts as games} \label{sect:es-contracts}

We assume a denumerable universe of \emph{events} $e, e', \ldots \in \Ev$, 
uniquely associated to 
\emph{participants} ${\pmv A}, {\pmv B}, \ldots \in \aname$
by a function $\princsym: \Ev \rightarrow \aname$.
For all ${\pmv A} \in \aname$, we write $\invprinc{\pmv A}$ for the set
$\setcomp{e \in \Ev}{\princ{e} = {\pmv A}}$.
For a sequence $\sigma = \seq{e_0 \, e_1 \cdots}$ in $E$ (possibly infinite), 
we write $\mkset{\sigma}$ for the set of elements in $\sigma$; 
we write  $\sigma_i$ for the subsequence $\seq{e_0 \cdots e_{i-1}}$
containing exactly $i$ events.
If $\sigma = \seq{e_0 \cdots e_n}$, 
we write $\sigma \, e$ for the sequence $\seq{e_0 \cdots e_n \, e}$.
% The predicate $\DF{\sigma}$ is true whenever $\sigma$ has no duplicates,
% i.e.\ for all $i \leq n$, $e_i \not\in \mkset{\sigma_i}$.
The empty sequence is denoted by~$\epsilon$.
For a set $S$, we denote with $S^*$ the set of finite 
sequences over $S$, and with $S^{\infty}$ the set
of finite and infinite sequences over $S$.

\smallskip
A contract is modelled in~\cite{BCZ13post} 
as a concurrent game
featuring \emph{obligations} (what I must do in a given state) 
and \emph{objectives} (what I wish to obtain).
Obligations are modelled as an event structure (ES).

\begin{definition}[{{\bf Event structure~\cite{Winskel86}}}]
  An event structure $\esname$ is a triple $\seq{E, \#, \vdash}$, where:
  \begin{itemize}
    
  \item $E$ is a set of events,
    
  \item $\# \;\subseteq E \times E $ is an irreflexive and symmetric 
    \emph{conflict} relation.
    For a set of events $X$, the predicate $\CF{X}$ is true iff $X$ is 
    \emph{conflict-free}, \text{i.e.} 
    $\CF{X} \triangleq ( \forall e,e' \in X: \neg (e \# e'))$.
    
  \item $\vdash \;\subseteq \setcomp{X \subseteqfin E}{\CF{X}} \; \times \; E$ 
    is the \emph{enabling} relation, which is \emph{saturated}, \text{i.e.}:
    \[
    \forall X \subseteq Y \subseteqfin E.\;\;
    X \vdash e  \,\land\, CF(Y) 
    \implies 
    Y \vdash e
    \]
  \end{itemize}
\end{definition}

Intuitively, an enabling $X \vdash e$ models the fact that, 
if all the events in $X$ have happened, 
then $e$ is an obligation for $\princ{e}$.
The conflict relation $\#$ is used to model
non-deterministic choices: if $e \# e'$ then $e$ and $e'$ cannot occur
in the same computation.
An obligation may be discharged only by performing the required event, 
or any event in conflict with it.
For instance, consider an internal choice between two events $e_a$ and~$e_b$.
This can be modelled by an \ES\ with enablings 
$\emptyset \vdash e_a$, $\emptyset \vdash e_b$ 
and conflict $e_a \# e_b$.
After the choice (say, of~$e_a$), the obligation $e_b$ is discharged.
The other component of a contract is
a function $\payoffsym$ which associates each participant {\pmv A} 
with a set of sequences in $\Ev^{\infty}$ 
(the set of finite or infinite sequences on $\Ev$), 
which enumerates all the executions
where {\pmv A} has a positive payoff.

\begin{definition}[\bf Contract] \label{def:es:contract}
A contract $\cname$ is a pair $\seq{\esname, \payoffsym}$, 
where:
\begin{enumerate}[(a)]

\item $\esname = \seq{E,\#,\vdash,\ell}$ is a \emph{labelled} event structure,
with $E \subseteq \Ev$ and labelling function $\ell: E \rightarrow \Act \cup \coAct$.

\item $\payoffsym: \aname \rightharpoonup \powset{E^{\infty}}$
associates each participant with a set of traces.

\end{enumerate}
\end{definition}

Note that $\payoffsym$ is a partial function 
(from $\aname$ to sets of event traces), 
hence a contract is not supposed to define payoffs for all the
participants in $\aname$.
Hereafter, we shall assume that if $\cname$ prescribes for {\pmv A}
some obligations, then $\cname$ must also declare {\pmv A}'s payoffs,
i.e.\ we ask that
$\payoff{(\princ{e})}{} \neq \bot$ whenever $X \vdash e$ in $\esname$.

Given two contracts $\cname,\cname'$, we denote with $\cname \mid \cname'$ 
their composition.
If $\cname$ is {\pmv A}'s contract and
$\cname'$ is the contract of an adversary {\pmv M} of {\pmv A}, 
then a na\"ive composition could easily lead to an attack,
e.g.\ {\pmv M}'s contract could say that {\pmv A} must
pay her 1M euros.
To avoid such kinds of attacks,
contract composition is a partial operation.
We do \emph{not} compose contracts which assign payoffs 
to the same participant.

\begin{definition}[\bf Contract composition] \label{def:es:ccomp}
We say that two contracts
$\cname = \seq{\esname,\payoffsym}$
and
$\cname' = \seq{\esname',\payoffsym'}$ 
are \emph{composable} iff
\(
  \forall {\pmv A} \in \aname.\;
  (\payoffsym({\pmv A}) = \bot \;\lor\;
  \payoffsym'({\pmv A}) = \bot)
\).
If $\cname$, $\cname'$ are composable, we define 
their composition as
\(
\cname \ccomp \cname' 
 =
\seq{\esname \sqcup \esname',\payoffsym \sqcup \payoffsym'}
\), where
$\esname \sqcup \esname' = \seq{E \cup E', \# \cup \#', \vdash \cup \vdash', \ell \cup \ell'}$.
\end{definition}

  % \paragraph{Agreements.}

A crucial notion on contracts is that of \emph{agreement}.
Intuitively, when {\pmv A} agrees on a contract $\cname$,
then she can safely initiate an interaction with the other participants, 
and be guaranteed that the interaction will not ``go wrong''
--- even in the presence of attackers.
This does not mean that {\pmv A} will always reach her objectives:
we intend that {\pmv A} agrees on a contract when, 
in all the interactions where she does not succeed, then 
some other participant must be found dishonest.
That is, we consider {\pmv A} satisfied if she can blame another participant.
In real-world applications, a judge may provide compensations to {\pmv A},
or impose a punishment to the participant who has violated the contract.
% Here, we shall not explicitly model the judge,
% and we shall focus instead on how to formalise the agreement property.

We interpret a contract $\cname = \seq{\esname,\payoffsym}$ 
as a multi-player game,
where the players concurrently perform events
in order to reach the objectives in $\payoffsym$.
A \emph{play} $\sigma$ of $\cname$ is a (finite or infinite) sequence 
of events of~$\esname$,
such that each event $e$ in $\sigma$ is enabled by its predecessors.
Formally, the plays of $\esname$ are the traces of
%the LTS 
the labelled transition system $\EvTS{\esname}$  induced by 
%We denote with $\ETS{\esname}$ the labelled transition systems induced by 
the relation 
% $\esname \etsrel{\ell(e)}{} \esname[e]$, 
% $\esname \dotarrow{\!\!e\!\!} \esname[e]$,
$\esname \xrightarrow{e} \esname[e],$
where $\emptyset \vdash e$, and $\esname[e]$ is the \emph{remainder} 
of $\esname$ after executing $e$. 
% (see App.~\ref{app:es}).

\begin{definition}[\bf Remainder of an ES]\label{def:es-remainder}
  For all ES $\esname = \seq{E, \#, \vdash, \ell}$ 
  and for all $e \in E$, we define
  the ES $\esname[e]$ as 
  $\seq{E', \#', \vdash', \ell'}$, where:
  \[
  \begin{array}{rcl}
    E'  & = &  E \setminus (\setenum{e} \cup \setcomp{e'}{e \# e'}) 
    \\[2pt]
    \#' & = &  \# \setminus (\setcomp{(e,e')}{e\#e'} \cup \setcomp{(e',e)}{e\#e'}) 
    \\[2pt]
    \vdash' & = &  \setcomp{(X \setminus \setenum{e} , e')}{(X,e')\in \;\vdash''} 
    \;\text{ where } \vdash'' \;=\;  \vdash \setminus \setcomp{(X , e')}{ e' \# e \,\lor\, e' = e \,\lor\, \neg \CF{ X \cup \setenum{e}} }
    \\[2pt]
    \ell' & = & \ell \setminus ( \setcomp{(e',\ell(e'))}{  e'\# e} \cup \setenum{(e,\ell(e)}) 
  \end{array}
  \]
\end{definition}

A \emph{strategy} $\Sigma$ for {\pmv A} is a function which associates 
to each finite play $\sigma$ a set of events of {\pmv A} 
(possibly empty), such that
if $e \in \Sigma(\sigma)$ then $\sigma e$ is still a play.
A play $\sigma = \seq{e_0 \, e_1 \cdots}$ \emph{conforms} to a strategy $\Sigma$ 
for {\pmv A} when, for all $i \geq 0$, 
$e_i \in \invprinc{\pmv A}$ implies $e_i \in \Sigma(\sigma_i)$.
A play is \emph{fair} w.r.t.\ a strategy $\Sigma$ iff any
event permanently prescribed by $\Sigma$ is eventually performed.

\begin{definition}[{{\bf Fair play}}] \label{def:es:fair-play}
A play $\sigma = \seq{e_0 \, e_1 \cdots}$ 
% (either finite or infinite) 
is \emph{fair} w.r.t.\ the strategy $\Sigma$ iff:
\[
  \forall i \leq |\sigma|.\;
  \big(
  \forall j : i \leq j \leq |\sigma|.\;
  e \in \Sigma(\sigma_j)
  \big)
  \implies 
  \exists h \geq i.\;
  e_h = e
\]
\end{definition}

% Our notion of agreement takes into account whether participants 
% behave honestly in their plays.
A participant {\pmv A} is \emph{innocent} in a play
if {\pmv A} has no persistently enabled events,
i.e.\ if all her enabled events are either performed or conflicted.
% (below, $\mkset{\sigma}$ denotes the set of events in $\sigma$,
% and $\sigma_i$ denotes the prefix of $\sigma$ containing exactly $i$ events).

\begin{definition}[{{\bf Innocence}}] \label{def:es:innocence}
We say {\pmv A} \emph{innocent in $\sigma$} iff
\(\;
  \forall i \geq 0.\;
  \forall e \in \invprinc{\pmv A}.\;
  (
  % \mkset{\sigma_i} \xrightarrow{\; e \;}_{\esname}
  \mkset{\sigma_i} \vdash e 
  \;\implies
  \exists j \geq i.\;
  % \mkset{\sigma_j} \not\xrightarrow{\;\;\; e}_{\esname}
  e_j \# e \;\lor\; e_j = e
  )
\).
If {\pmv A} is not innocent in $\sigma$, then we say she is \emph{culpable}.
\end{definition}

We now define when a participant \emph{wins} in a play.
If {\pmv A} is culpable, then she loses.
If {\pmv A} is innocent, but some other participant is culpable,
then {\pmv A} wins.
Otherwise, if all participants are innocent, then 
{\pmv A} wins if she has a positive payoff in the play.
Formally, $\sigma$ is a winning play of {\pmv A} iff 
$\sigma \in \win{\pmv A}{}$, defined below.

\begin{definition}[\bf Winning plays and strategies] \label{def:es:win}
We define the function $\winsym: \aname \rightarrow \powset{\Ev^{\infty}}$ 
as follows:
\begin{align*}
  \win{\pmv A}{} 
  \; = \;
  & \setcomp{\sigma \in \payoff{\pmv A}{}}
  {\forall {\pmv B} : {\pmv B} \text{ innocent in } \sigma}
  \; \cup
  \setcomp{\sigma}{\text{{\pmv A} innocent in $\sigma$, and $\exists {\pmv B} \neq {\pmv A} : {\pmv B} \text{ culpable in } \sigma$}}
\end{align*}
We say that $\Sigma$ is \emph{winning} for {\pmv A} in $\cname$
iff {\pmv A} wins in every fair play of $\cname$ which conforms to $\Sigma$.
\end{definition}

% We now define when a participant \emph{agrees} on a contract.
Intuitively, {\pmv A} agrees with $\cname$ when she has a strategy $\Sigma$
which allows her to win in all fair plays conform to~$\Sigma$.
% When this holds, we say that {\pmv A} agrees on $\cname$.
Note that neglecting unfair plays is quite reasonable:
indeed, an unfair scheduler could easily prevent 
an honest participant (ready to fulfil all her obligations)
from performing any action.

\begin{definition}[\bf Agreement] \label{def:es:agreement}
A participant {\pmv A} \emph{agrees} on $\cname$ whenever
{\pmv A} has a winning strategy in $\cname$.
% bart: "admits an agreement" non serve in questo lavoro
% and that $\cname$ \emph{admits an agreement} whenever 
% all the involved participants agree on $\cname$.
\end{definition}

\section{Compliance as agreement}
  \begin{figure}[t]
  \hrulefill
  \[
  \small
  \begin{array}{rcl}
    \sem[\pmv A]{\success}{\rho} & = & \seq{\setenum{e}, \emptyset, \setenum{(\emptyset,e)}, \setenum{(e,\checkmark)}}
    \;\text{ where } e\in\invprinc{\pmv A}
    \\[4pt]
    \sem[\pmv A]{x}{\rho} & = & \rho(x)\ =\ \seq{E, \#, \vdash, \ell}\ \;\text{where}\ E\subseteq \invprinc{\pmv A} 
    \\[4pt]
    \sem[\pmv A]{\alpha.P}{\rho} & = & 
    \seq{E \cup \setenum{e_{\alpha}},\; \#,\; \vdash',\; \ell  \cup \setenum{(e_{\alpha}, \alpha)}} 
    \;\text{ where }
    \sem[\pmv A]{P}{\rho}  =  \seq{E,\; \#,\; \vdash, \; \ell},\; 
    e_{\alpha} \in \invprinc{\pmv A}\setminus E, \;\text{ and }
    \\
    &&  \vdash' \;=\;  \setenum{(\emptyset,e_{\alpha})} \cup \setcomp{(\setenum{e_{\alpha}}\cup X,e)}{(X,e) \in \;\vdash}
    \\[4pt]
    \sem[\pmv A]{\bigodot_{i \in I}{P_i}}{\rho} & = & \seq{\bigcup E_i,\; \#, \; \bigcup\vdash_i,\; \bigcup \ell_i} 
    \text{ where}\ 
    \sem[\pmv A]{P_i}{\rho} =  \seq{E_i,\; \#_i,\; \vdash_i, \; \ell_i},\; E_i \;\text{ pairwise disjoint, and } \\
    & & \#\ =\  \bigcup \#_i \; \cup \; \setcomp{(e,e')}{  (\emptyset, e) \in \;\vdash_i \;\land\; (\emptyset, e') \in \;\vdash_j \; \land \;  i \neq j},
    \;\text{ with }\ \bigodot \in\setenum{\sum, \bigoplus} 
    \\[4pt]
    \sem[\pmv A]{\rec{x}{P}}{\rho} & = & \mathit{fix}\ \Gamma \;\text{ where}\ \Gamma(\esname) = \sem[\pmv A]{P}{\rho\setenum{\bind{x}{\esname}}}
    \\[4pt] 
    \sem[\pmv A_1 \pmv A_2]{P_1 \mmid P_2}{\rho} \hspace{-11pt} & \;= & 
    \seq{E,\; \#_1\cup \#_2,\; \vdash,\; \ell}\ \text{where}\ \sem[\pmv A_i]{P_i}{\rho} = \seq{E_i,\; \#_i,\; \vdash_i,\; \ell_i},
    \;
    E  =  E_1 \cup E_2\ \text{with}\ E_1 \cap E_2 = \emptyset,\  
    \ell  =  \ell_1\cup \ell_2, \\
    & & \vdash\ =\   \setcomp{(X \cup Y, e)}{\ell(e) \in \coAct \land (X, e) \in \vdash_i \land 
      \forall e' \in X. \exists e'' \in Y. e''\in E\setminus E_i \land \ell(e') = \overline{\ell(e'')}} 
    \\[0pt]
    & &       \hspace{9pt}\cup\ \;\,\{(X \cup Y \cup \setenum{\hat{e}}, e)\ \mid\ \ell(e) \in \Act \;\land\; (X, e) \in \;\vdash_i \;\land\; 
    \forall e' \in X.\; \exists e'' \in Y.\; \big( \\
    & &       \hspace{101pt}  e''\in E\setminus E_i \land\; \ell(e') = \overline{\ell(e'')} 
    \;\land\; \hat{e} \in E\setminus E_i 
    \;\land\; \ell(\hat{e}) = \overline{\ell(e)}\big)\}
    \\[-5pt]
  \end{array}
  \] 
  \hrulefill
  \vspace{-5pt}
  \caption{Denotational semantics of session types.}\label{fig:es-denotational}
\end{figure}

  %%%%%%%%%%%%%%%%%%%%%%%%%%%%%%%%%%%%%%%%%%%%%%%%%%%%%%%%%%%%%%%%%%%%%%%%%%%%%
%%%                                                                       %%%
%%%                                   st-to-es                            %%%
%%%                                                                       %%%
%%%%%%%%%%%%%%%%%%%%%%%%%%%%%%%%%%%%%%%%%%%%%%%%%%%%%%%%%%%%%%%%%%%%%%%%%%%%%

We now relate session types with contracts.
To do that, we start by introducing an event structure semantics 
for session types.
This denotational semantics is then related to a 
turn-based operational semantics of session types
(\Cref{fig:sb-contracts:turn-semantics}),
which preserves the notion of compliance (\Cref{lem:sb-event-compliance}).
In~\Cref{def:sb-to-contract} we transform 
session types into contracts.
\Cref{th:compliance-iff-eager-winning} 
establishes a correspondence
between compliance of session types and 
winning strategies in contracts. 

\begin{definition}[\bf ES semantics of session types]
\label{def:es-denotational}
The denotation of session types is defined by the rules in 
\Cref{fig:es-denotational},
where $\rho$ is an environment mapping variables $x$ to ESs.
\end{definition}

The denotation of session types is almost straightforward. 
Note that the parameter {\pmv A} is used to associate all the events 
of the constructed ES to that participant. 
The enabling relation of compositions of session types 
takes into account the different 
flavour of the events (actions) involved. 
Intuitively, an action $\coact{b}$ in a contract such as $\coact{a}.\coact{b}$ 
must wait for its prefix $\coact{a}$, and for a matching  $\act{a}$-synchronized action. 
On the other hand, an action in $\Act$ such as $\act{c}$ 
must also wait to be matched by a synchronizing action $\coact{c}$. 
This behaviour is simulated in the event structure of the contract: 
for an enabling $X \vdash e$ with $\ell(e) \in \coAct$, 
we add to $X$ the set $Y$ of all the matching events of $X$. 
Instead, for an enabling $X \vdash e’$ with $\ell(e’) \in \Act$, 
we also add to $X$ and $Y$ the coaction of $\ell(\act{e’})$.

As session types have recursion, the standard machinery on fixed points is needed. 
Henceforth, following closely what is done in~\cite{Winskel86}, 
we introduce a notion of partial ordering 
on event structures. 
The intuition is that $\esname$ is less or equal to $\esname'$ 
whenever each configuration of the former is a configuration of the latter, 
and each configuration of $\esname'$ where the events are those 
of $\esname$ is a configuration of 
$\esname$ as well. 
% Formally:

\begin{definition}[\bf Ordering of ESs]\label{def:es-ordering}
  Let $\esname  = \seq{E,\#,\vdash,\ell}$ 
  and $\esname'  = \seq{E',\#',\vdash',\ell'}$ be two ESs.
  Then we write $\esname\ \trianglelefteq\ \esname'$ iff:
  \begin{itemize}
    
  \item $E\subseteq E'$, $\# \subseteq \#'$, $\vdash\ \subseteq\ \vdash'$ and $\forall e\in E$. $\ell'(e) = \ell(e)$,  
    
  \item for all $e_1, e_2\in E$, if $(e_1,e_2) \in \#'$ 
    then $(e_1,e_2) \in \#$, and 
    
  \item for all $X\subseteq E$, $e\in E$, 
    if $(X,e)\in \;\vdash'$ then $(X,e)\in \;\vdash$.
    
  \end{itemize}
\end{definition}

The relation $\trianglelefteq$ is a partial order on event structures. 
An $\omega$-chain of ESs 
$\esname_1 \trianglelefteq \esname_2 \trianglelefteq\dots 
\trianglelefteq\esname_n\trianglelefteq  \dots$ has a least 
upper bound defined as 
$\bigsqcup \esname_i = (\bigcup_i E_i, \bigcup_i \#_i, \bigcup_i \vdash_i, 
\bigcup_i \ell_i)$. 
The ES 
$\underline{\emptyset} = \seq{\emptyset, \emptyset, \emptyset, \emptyset}$ 
is the least element of the partial order. 
Given a unary operator $\mathbf{F}$ on event structures, we say that it is 
\emph{continuous on events} iff for every $\omega$-chain of ESs 
$\esname_1 \trianglelefteq \esname_2 \trianglelefteq\dots 
\trianglelefteq\esname_n\trianglelefteq  \dots$ it holds that 
$\mathbf{F}(\bigcup_i E_i) = \bigcup_i \mathbf{F}(E_i)$. 
If furthermore the operator $\mathbf{F}$ is monotonic 
with respect to $\trianglelefteq$ then $\mathbf{F}$ is \emph{continuous}. 
Given a continuous unary operator $\mathbf{F}$, we can then 
define its fixed point standardly using Tarski's theorem, 
as event structures with $\trianglelefteq$ are a complete partial order 
with bottom. 
The fixed point is denoted by 
$\mathit{fix}\ \Gamma = \bigsqcup \mathbf{F}(\underline{\emptyset})$.
It is standard to prove that the operators defined by the denotational semantics 
in~\Cref{fig:es-denotational} are continuous.
  %%%%%%%%%%%%%%%%%%%%%%%%%%%%%%%%%%%%%%%%%%%%%%%%%%%%%%%%%%%%%%%%%%%%%%%%%%%%%
%%%                                                                       %%%
%%%                        compliance-to-agreement                        %%%
%%%                                                                       %%%
%%%%%%%%%%%%%%%%%%%%%%%%%%%%%%%%%%%%%%%%%%%%%%%%%%%%%%%%%%%%%%%%%%%%%%%%%%%%%

\medskip
We shall now relate the denotational semantics in~\Cref{def:es-denotational} with 
an operational semantics of binary session types
where the two participants alternate in firing actions 
(\Cref{fig:sb-contracts:turn-semantics}).
To do that, we extend the syntax of session types
with the term $[\coact{a}]P$, where $[\coact{a}]$ models a one-position buffer 
storing~$\coact{a}$.
Also, we tacitly assume 
% the syntax to be up-to 
unfolding of recursion.
A participant with an internal choice $\coact{a}.P \sumInt Q$ 
can fire the action $\coact{a}$ (if the buffer is empty),
and write $\coact{a}$ to the buffer.
The next turn is of the other participant, which can empty the buffer
by firing $\act{a}$ in an external choice.
To be coherent with the event structure semantics, 
we also assume that the success state $\success$ fires an action
$\checkmark \in \coAct$ before reaching the stuck state $\nil$.

The following theorem relates the denotational and the turn-based
operational semantics of session types.
Their (action-labelled) LTSs are strongly bisimilar.
Below, we denote with $\ETS{\esname}$ the transition systems induced by 
the relation $\esname \xrightarrow{e} \esname'$, 
by relabelling transitions with actions $\ell(e)$,
and we denote with $\TS{P}$ the labelled transition system induced by
the turn-based relation~$\cmove{}$.

\begin{theorem} \label{th:sb-to-es}
For all session types $P,Q$, we have
$\TS{P\mmid Q} \sim \ETS{\sem{P\mmid Q}{}\!}$.
\end{theorem}

The turn-based semantics of session types preserves 
the compliance relation of~\Cref{def:compliance}.

\begin{lemma} \label{lem:sb-event-compliance}
$P \dashv Q$ iff
$P \mmid Q \cmove{}^* P' \mmid Q' \not\cmove{}\;$ implies $P' = \nil$.
\end{lemma}

\begin{figure}[t]
% \vspace{10pt}
\hrulefill
\vspace{-10pt}
\small
\[
\begin{array}{c}
  (\sumI{\coact{a}}{P} \,\sumInt\, Q)
  \;\mmid\;
  R
  \;\cmove{\coact{a}}\;
  [\coact{a}]{P}
  \;\mmid\;
  R
  \hspace{40pt}  
  (\sumE{\act{a}}{P} \,\sumExt\, Q)
  \;\mmid\;
  [\coact{a}]{R}  
  \;\cmove{\act{a}}\;
  P \mmid R
  % \rec X P \equiv P\setenum{\bind{X}{\rec X P}}
  \hspace{40pt}
  \success \mmid P \;\cmove{\checkmark}\; \nil \mmid P
  \\[-10pt]
\end{array}
\]
\hrulefill
\vspace{-10pt}
\caption{Turn-based operational semantics of session types
(symmetric rules omitted).}
\label{fig:sb-contracts:turn-semantics}
\end{figure}

We now define a transformation from session types $P$ to contracts,
denoted by $\cname_{\pmv A}(P)$.
The parameter {\pmv A} is used to properly assign 
the obligations and the objective to participant {\pmv A}.

\begin{definition}[\bf Contract of a session type]
\label{def:sb-to-contract}
For all session types $P$ and participants {\pmv A}, we define the
contract $\cname_{\pmv A}(P)$ as $\seq{\sem[\pmv
  A]{P}{\emptyset},\payoffsym}$, where \( \payoff{\pmv A} =
\setcomp{\sigma \in \Ev^{\infty}}{\sigma \in \Ev^* \implies \exists e
  \in \mkset{\sigma} \cap  \invprinc{A}: \ell(e) = \checkmark}
\).
% (E^{\infty} \setminus E^*)\cup \setcomp{\sigma \in E^*}{\exists e \in \mkset{\sigma} : \ell(e) = \success}$
\end{definition}

We now establish a correspondence between compliance in session types
and the existence of certain winning strategies in contracts. 
To do that, we consider strategies which ensure {\pmv A} to be
innocent in every (fair) play.
The greatest of such strategies is the \emph{eager strategy}
\(
  \Sigma_{\pmv A}(\sigma)
  =
%  \lambda \sigma.\,
  \setcomp{e \in \invprinc{\pmv A}}{\mkset{\sigma} \vdash e \;}
\)
which prescribes {\pmv A} to do all her enabled events.
The session type $P$ (say, of participant {\pmv A}) is compliant with $Q$ 
iff the eager strategy is winning for {\pmv A} in the contract 
$\cname_{\pmv A}(P) \mid \cname_{\pmv B}(Q)$.

\begin{theorem} \label{th:compliance-iff-eager-winning}
$P \dashv Q$
iff the eager strategy
is winning for {\pmv A} in $\cname_{\pmv A}(P) \mid \cname_{\pmv B}(Q)$.
\end{theorem}

By the theorem above, it follows that compliance implies agreement.

\begin{corollary} \label{cor:compliance-to-agreement}
If $P \dashv Q$, then
{\pmv A} agrees on $\cname_{\pmv A}(P) \mid \cname_{\pmv B}(Q)$.
\end{corollary}

Note that the converse implication does \emph{not} hold:
for instance, 
for $P = \coact{a}.\coact{c} \sumInt \coact{b}$
and $Q = \act{a} \sumExt \act{b}$,
we have that $P \not\dashv Q$, 
but {\pmv A} agrees on $\cname_{\pmv A}(P) \mid \cname_{\pmv B}(Q)$.
Indeed, choosing the branch $\coact{b}$ 
leads to a winning strategy for~{\pmv A}.
Note instead that $P$ is \emph{not} weakly compliant with $Q$
according to~\cite{Padovani10tcs},
because no orchestrator can prevent {\pmv A} from
choosing the branch~$\coact{a}$.
However, $P' = \coact{a}.\coact{c} \sumExt \coact{b}$
is weakly compliant with $Q$, because the orchestrator
can resolve the external non-determinism by choosing the branch $\bar{b}$.
Weak compliance can be formalised in game-based contracts
by modelling the orchestrator as a third player of the game
(who can use \emph{any} strategy to favour the interaction between {\pmv A} and {\pmv B}),
and by adapting the construction of the contracts to take into
account for the moves of the orchestrator.

  %%%%%%%%%%%%%%%%%%%%%%%%%%%%%%%%%%%%%%%%%%%%%%%%%%%%%%%%%%%%%%%%%%%%%%%%%%%%%
%%%                                                                       %%%
%%%                                   example                             %%%
%%%                                                                       %%%
%%%%%%%%%%%%%%%%%%%%%%%%%%%%%%%%%%%%%%%%%%%%%%%%%%%%%%%%%%%%%%%%%%%%%%%%%%%%%

\paragraph{An example.}
We now illustrate with the help of an example the transformation 
from session types to game-based contracts.
Below,  we use the following shorthands:
% $X \vqdash Y$ for $\forall e \in Y.\; X \vdash e$,
$a \vdash b$ for $\setenum{a} \vdash b$,
and $\vdash e$ for $\emptyset \vdash e$.

Consider two participants $\pmv{A}$ and $\pmv{B}$, with session types 
$P = \coact{a} \sumInt \coact{b}.\coact{a} $ and 
$Q = \act{a}.\act{b} \sumExt \act{b}.\act{a} \sumExt \act{c}$,
respectively
(trailing $\success$s are omitted). 
According to~\Cref{def:compliance},
the session type of $\pmv{A}$ is compliant with that of $\pmv{B}$, 
while the converse does not hold. 
Below we construct the event structures associated to $P$ and $Q$, 
and the one associated to the their composition $P\mmid Q$.
To ease the reading, we decorate actions in $P$ and $Q$ with the events
they will be associated with in the event structures;
we stipulate that the events of $\pmv{A}$ have odd indexes,
whereas those of $\pmv{B}$ have even ones.
Hence, we have:
\[
P =
\coact{a}_{e_1}.{\success}_{e_3} \sumInt
\coact{b}_{e_5}.\coact{a}_{e_7}.{\success}_{e_9} \quad \text{ and } \quad 
Q =
\act{a}_{e_2}.\act{b}_{e_4}.{\success}_{e_6} \sumExt
\act{b}_{e_8}.\act{a}_{e_{10}}.{\success}_{e_{12}} \sumExt
\act{c}_{e_{14}}.{\success}_{e_{16}} 
\]

\noindent
By the construction in Def.~\ref{def:es-denotational}, we have:
\[
\sem[\pmv A]{P}{\rho} = (\setenum{e_1,e_3,e_5,e_7,e_9}, \quad
\setenum{ e_1 \# e_5}, \quad \setenum{ \vdash e_1, \vdash e_5, e_1
  \vdash e_3, e_5 \vdash e_7, e_7 \vdash e_9 }, \quad \ell_\pmv{P} )
\]  
where
$\ell_\pmv{P}(e_1) = \coact{a}$, $\ell_\pmv{P}(e_5) = \coact{b}$, 
$\ell_\pmv{P}(e_7) = \coact{a}$, and the others are labelled with $\checkmark$. 
Furthermore:
\[
\sem[\pmv B]{Q}{\rho} = 
(\setenum{e_2,e_4,e_6,e_8,e_{10}, e_{12},e_{14}, e_{16}}, 
\hspace{2pt} 
\setenum{ e_2 \# e_8, e_2 \# e_{14}, e_8 \# e_{14}}, 
\hspace{2pt} \{
\begin{array}{l}
  \vdash e_2, \vdash e_8, \vdash  e_{14}, e_2  \vdash e_4, e_4 \vdash e_6 \\
  e_8 \vdash e_{10}, e_{10} \vdash e_{12}, e_{14} \vdash e_{16} 
\end{array}
\}, \hspace{2pt} 
\ell_\pmv{Q} )
\]
where $\ell_\pmv{Q}(e_2) = \act{a} = \ell_\pmv{Q}(e_{10})$,   
$\ell_\pmv{Q}(e_4) = \act{b} = \ell_\pmv{Q}(e_{8})$, $\ell_\pmv{Q}(e_{14}) = \act{c}$,
and the other events are labelled with $\checkmark$.
The event structure associated to $P \mmid Q$ is: 
\[
\sem[\pmv A, \pmv B]{P \mmid Q}{\emptyset} = (E_P \cup E_Q,
\quad \#_P \cup \#_Q, \quad \vdash, \quad \ell_P \cup \ell_Q )
\tag*{where:}
\]
\[ 
\vdash \quad  = \quad 
\begin{array}{l}
  \vdash e_1, \vdash e_5,  \setenum{e_1,e_2} \vdash e_3, \setenum{e_1,e_{10}} \vdash e_3,
  \setenum{e_5,e_8} \vdash e_7, \setenum{e_5,e_4} \vdash e_7, \setenum{e_2,e_7} \vdash e_9, 
  \setenum{e_7,e_{10}} \vdash e_9, \\
  e_1\vdash e_2, e_7 \vdash e_2, \setenum{e_1,e_2,e_5} \vdash e_4, \setenum{e_7,e_2,e_5} \vdash e_4, 
  \setenum{e_4,e_5}\vdash e_6, e_5 \vdash e_8, 
  \setenum{e_8, e_5, e_1} \vdash e_{10}, \\ \setenum{e_8, e_5, e_7} \vdash e_{10},
  \setenum{e_{10},e_1}\vdash e_{12},\setenum{e_{10},e_7}\vdash e_{12}
\end{array}
\]
The event-labelled transition system of $\sem{P\mmid Q}{}$ 
and the eager strategy $\Sigma_{\pmv A}$ of {\pmv A}
are depicted below:
\vspace{-10pt}
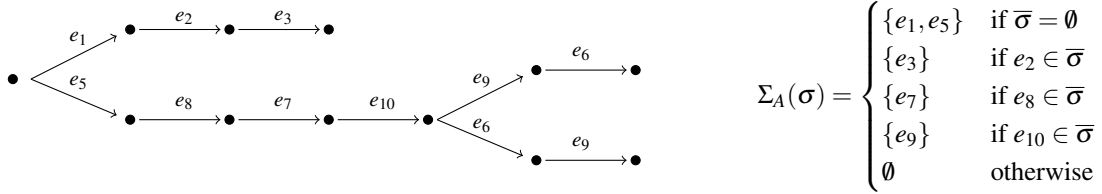
\begin{center}
  \begin{tabular}{cc}
    \begin{minipage}[!b]{0.65\textwidth}
      \[
      \begin{tikzpicture}[scale=1.2]
        \draw [fill=black] (-0.2 ,0.2) circle (0.05);
        % branch 1.
        \draw [->] (0,0.2)  -- (0.95,0.68) ;
        \node [above] at (0.53, 0.47)  {{\scriptsize {$e_1$}}};
        \draw [fill=black] (1.1 ,0.75) circle (0.05);
        \draw [->] (1.2,0.75)  -- (2.1,0.75) ;
        \node [above] at (1.7, 0.7)  {{\scriptsize {$e_2$}}};
        \draw [fill=black] (2.2 ,0.75) circle (0.05);
        \draw [->] (2.3,0.75)  -- (3.2,0.75) ;
        \node [above] at (2.8, 0.7)  {{\scriptsize {$e_3$}}};
        \draw [fill=black] (3.3 ,0.75) circle (0.05);
        % 
        % branch 2.
        \draw [->] (0,0.2)  -- (0.95,-0.2) ;
        \node [above] at (0.53, -0.02) {{\scriptsize{$e_5$}}}; 
        \draw [fill=black] (1.1 ,-0.25) circle (0.05); 
        \draw [->] (1.2,-0.25)  -- (2.1,-0.25) ;
        \node [above] at (1.7, -0.25)  {{\scriptsize {$e_8$}}};
        \draw [fill=black] (2.2 ,-0.25) circle (0.05);
        \draw [->] (2.3,-0.25)  -- (3.2, -0.25) ;
        \node [above] at (2.8, -0.25)  {{\scriptsize {$e_7$}}};
        \draw [fill=black] (3.3 ,-0.25) circle (0.05); 
        \draw [->] (3.4,-0.25)  -- (4.3, -0.25) ;
        \node [above] at (3.9, -0.25)  {{\scriptsize {$e_{10}$}}};
        \draw [fill=black] (4.4 ,-0.25) circle (0.05);
        % 
        % branch 2.1  
        \draw [->] (4.5,-0.25)  -- (5.45,0.23) ;
        \node [above] at (5, -0)  {{\scriptsize {$e_9$}}};  
        \draw [fill=black] (5.6 ,0.3) circle (0.05);
        \draw [->] (5.7,0.3)  -- (6.6,0.3) ; 
        \node [above] at (6.1, 0.3)  {{\scriptsize {$e_6$}}};  
        \draw [fill=black] (6.7 ,0.3) circle (0.05);
        % branch 2.1   
        \draw [->] (4.5,-0.25)  -- (5.45,-0.65) ;
        \node [above] at (5, -0.5)  {{\scriptsize {$e_6$}}}; 
        \draw [fill=black] (5.6 ,-0.7) circle (0.05); 
        \draw [->] (5.7,-0.7)  -- (6.6,-0.7) ; 
        \node [above] at (6.1, -0.7)  {{\scriptsize {$e_9$}}};  
        \draw [fill=black] (6.7 ,-0.7) circle (0.05);
      \end{tikzpicture}
      \]
    \end{minipage}
    &
    \begin{minipage}[!t]{0.3\textwidth}
      \small
      \[
      \Sigma_{\pmv A}(\sigma) = \begin{cases} 
        \setenum{e_1, e_5} & \text{if } \mkset{\sigma} = \emptyset \\
        \setenum{e_3}      & \text{if } e_2 \in \mkset{\sigma} \\ 
        \setenum{e_7}      & \text{if } e_8 \in \mkset{\sigma} \\
        \setenum{e_9}      & \text{if } e_{10} \in \mkset{\sigma} \\  
        \emptyset & \text{otherwise}
      \end{cases}
      \]
    \end{minipage}
  \end{tabular}
\end{center}

\smallskip
We can see that {\pmv A} wins 
% (the events $e_3$ and $e_9$ are labelled with $\checkmark$)  
in all the fair plays which conform to the eager strategy $\Sigma_{\pmv A}$.
Since $\Sigma_{\pmv A}$ is winning, then {\pmv A} agrees on $\cname_{\pmv A}(P) \mid \cname_{\pmv B}(Q)$. 
Then, by~\Cref{th:compliance-iff-eager-winning}, $P \dashv Q$.

On the contrary, we notice that \pmv{B} has no winning strategies: 
indeed, whenever {\pmv A} chooses to perform event $e_1$,
then \pmv{B} is obliged to fire $e_2$ to recover his innocence, and 
then he gets stuck (and non-successful) when {\pmv A} fires $e_3$.
Then, by~\Cref{th:compliance-iff-eager-winning} 
it follows that $Q \not\dashv P$.

  \section{Conclusions}

We have related the notion of compliance in binary session types with
the one of agreement in game-based contracts.  
In particular, we have
shown that two session types are compliant if and only if their
encodings in game-based contract admit an agreement via a winning
\emph{eager} strategy (\Cref{th:compliance-iff-eager-winning}).

A relevant question is whether non-eager strategies are meaningful to
define weaker notions of compliance for session types.
This mostly depends on the interpretation of the internal choice operator $\sumInt$.
The usual meaning of an internal choice $\coact{a} \sumInt \coact{b}$ of
a participant {\pmv A} is that {\pmv A} is willing to opt between the two choices,
and \emph{both} of them must be available as external choices of the
other participant {\pmv B}.

Just to give a more realistic flavour to our scenario, 
assume that {\pmv B} is a bartender which only accepts payments in cash,
while {\pmv A} is a customer willing to pay either by cash or by credit card.
Under the standard notion of compliance, the two session types:
\[
  P_{\pmv A} \; = \; \coact{payCash} \sumInt \coact{payCC}
  \hspace{50pt}
  P_{\pmv B} \; = \; \act{payCash}
\]
are \emph{not} compliant, and so (by~\Cref{th:compliance-iff-eager-winning})
the eager strategy is \emph{not} winning in 
$\cname_{\pmv A}(P_{\pmv A}) \mid \cname_{\pmv B}(P_{\pmv B})$.

A different interpretation of the internal choice of {\pmv A} would be the following:
{\pmv A} is willing to choose between $\coact{payCash}$ and $\coact{payCC}$
if both options are available,
but she will also accept to pay cash (resp.\, to pay by credit card) 
if this is the only option available.
This interpretation is coherent with the fact that the contract
$\cname_{\pmv A}(P_{\pmv A}) \mid \cname_{\pmv B}(P_{\pmv B})$
admits an agreement,
via a non-eager strategy which requires {\pmv A} to renounce to the 
$\coact{payCC}$ alternative.

Similarly, we expect that other interpretations of compliance for session types
(e.g.\ that in~\cite{Castagna09toplas,BTZ12coordination}, 
where internal \emph{vs.} internal choices
and external \emph{vs.} external choices may be compliant, in some cases)
can be related to game-based agreements, via suitable (sub)classes of strategies.

  {\small \paragraph{Acknowledgments.}
This work has been partially supported by
Aut.\ Reg.\ of Sardinia grants L.R.7/2007
CRP-17285 (TRICS) and P.I.A.\ 2010 (``Social Glue''),
by MIUR PRIN 2010-11 project ``Security Horizons'',
and by EU COST Action IC1201
``Behavioural Types for Reliable Large-Scale Software Systems''
(BETTY).}

\bibliographystyle{eptcs}
\bibliography{main}

\iftoggle{proofs}{%
\begin{appendices}
\input{es.tex}
\end{appendices}
}
{}

\end{document}